\documentclass[aps, epsf, amsfonts]{iopart}
\usepackage{graphics}
\usepackage{graphicx}
\usepackage{epsfig}

\def\beq{\begin{equation}}
\def\eeq{\end{equation}}
\def\bea{\begin{eqnarray}}
\def\eea{\end{eqnarray}}
\def\hp{h_{+}}
\def\hc{h_{\times}}

\def\fp{F^{+}}
\def\fc{F^{\times}}

\begin{document}
\input epsf.tex

\title{The Effect of Data Gaps on LISA Galactic Binary Parameter Estimation.}
\author{J\'er\^ome Carr\'e \& Edward K. Porter}
\address{APC, UMR 7164, Universit\'e Paris 7 Denis Diderot,\\
		10, rue Alice Domon et L\'eonie Duquet, 75205 Paris Cedex 13, France}
\ead{porter@apc.univ-paris7.fr \\ carre@apc.univ-paris7.fr}
\vspace{1cm}
\begin{abstract}
\noindent In the last few years there has been an enormous effort in parameter estimation studies for different sources with the space based gravitational
wave detector, LISA.  While these studies have investigated sources of differing complexity, the one thing they all have in common is they assume continuous
data streams.  In reality, the LISA data stream will contain gaps from such possible events such as repointing of the satellite antennae, to discharging static charge
build up on the satellites, to disruptions due to micro-meteor strikes.  In this work we conduct a large scale Monte Carlo parameter estimation simulation for 
galactic binaries assuming data streams containing gaps.  As the expected duration and frequency of the gaps are currently unknown, we have decided to focus on gaps
of approximately one hour, occurring either once per day or once per week.  We also study the case where, as well as the expected periodic gaps, we have a 
data drop-out of one continuous week.  Our results show that for for galactic binaries, a gap of once per week introduces a bias of between 0.5\% and 1\% in the estimation of parameters,  for the most important parameters such as the sky position, amplitude and frequency.  This number rises to between 3\% and 7\% for the case of one gap a day, and to between 4\% and 9\% when we have one gap a day and a spurious gap of a week.    A future study will investigate the effect of data gaps on supermassive black hole binaries and extreme mass ratio inspirals.
\end{abstract}

\maketitle

\section{Introduction}

The planned ESA-NASA Laser Interferometer Space Antenna (LISA)~\cite{Danzmann,Bender} will be the first space-based gravitational wave (GW) detector. It will consist of three identical free-falling spacecraft in an equilatoral triangle configuration, with each spacecraft separated by $5\times 10^9$ m.  The constellation orbits around the Sun, $20^o$ behind the Earth, inclined at an angle of $60^o$ to the ecliptic. It is expected that LISA will detect sources in the frequency range of $10^{-5}-10^{-1}$ Hz. This should allow us to detect and extract parameters for such sources such as galactic white dwarf binaries (GBs), supermassive black hole binaries (SMBHBs) to redshifts of $z\approx20$~\cite{hughes,pc}, extreme mass ratio inspirals (EMRIs) to approximately $z=1$~\cite{bc,gair}, kinks and cusps of cosmic superstrings and a stochastic background from the background of low mass black hole binaries~\cite{mldc3}. 

In general, the sources detected by LISA will have a large signal to noise ratio (SNR).  As the sources will be loud, an ideal method to detect and extract the parameters from the various sources is matched filtering~\cite{Helstrom}. This technique works by cross correlating the received signal output with theoretical waveform models or templates. As matched filtering is especially dependent on the phase information of the waveform, a high correlation between signal and template then allows us to make a prediction of the source parameters.  

A number of studies have already been conducted on parameter estimation for a range of different sources. While these works have focused on different source types, the one thing they have in common is, each study assumes uninterrupted data.   In reality we know that communication between the spacecraft and Earth will be interrupted at various times during the mission. The interruptions are likely to come from such as events as antenna re-alignment, discharging any static charge build-up on the spacecraft, micro-meteor impacts or hardware failures either on the spacecraft or on Earth. The gaps in the data not only cause a loss of signal, but if not properly treated could generate, due to the discontinuity in the time domain, massive power leakage in the Fourier domain where the data analysis is conducted.  This excess power would result in spurious detection statistics and parameter estimation. 

While a number of other works have focused on the implementation of time domain tapers for the termination of inspiral waveforms where we are not taking the merger and ringdown into account~\cite{pc, sathya}, to our knowledge, there are only a handful of works that have investigated the influence of data gaps on GW parameter estimation~\cite{sylvestre,pollack}.   In this paper we propose a first-step attempt to deal with the treatment of data gaps that are irregular in both duration and frequency,  and investigate the effects on parameter estimation for white dwarf - white dwarf galactic binary systems.

The paper is organized as follows. In Section \ref{LISA+gaps} we present the response of the LISA detector in the low frequency approximation, the polarizations of the GWs and details how we treat the data gaps. In Section \ref{Monte Carlo} we explain the setup of the Monte Carlo (MC) simulation and outline some of the terminology related to GW data analysis. Finally, in Section \ref{Results} we present the results of our simulations.


\section{The LISA response and the treatment of gaps}
\label{LISA+gaps}
In this study, we chose to model the LISA response to an incoming GW in the low frequency approximation (LFA)~\cite{Cutler}.  It has been
shown that the LFA provides a realistic model of the LISA response up to a GW frequency of 5 mHz~\cite{CornishRubbo}. 

\subsection{The LISA response in the low frequency approximation}
The LFA response of a single LISA channel to an impending gravitational wave with both polarizations is
\beq
h(t) = \hp(t)\,\fp(t) + \hc(t)\,\fc(t),
\eeq
where the two polarizations of the gravitational wave $h_{+,\times}(t)$ are given for a non-chirping galactic binary by.  
\beq
h_{+}\left(t\right) = A_{0}\left(1+\cos^{2}\iota\right)\cos\left(\Phi\left(t\right)+\varphi_{0}\right), 
\eeq
\beq
h_{\times}\left(t\right) = -2\,A_{0}\cos\iota\,\sin\left(\Phi\left(t\right)+\varphi_{0}\right),
\eeq
where $A_{0}$ is a constant amplitude, $\iota$ is the angle of inclination of the source's orbital plane with respect to the observer, and $\varphi_{0}$ is a constant initial phase.  The time dependent phase of the gravitational wave, $\Phi(t)$, for a circular orbit is defined by
\beq
\Phi(t) = 2\pi f_{0}\left(t + R_{\oplus}\,\sin\theta\cos\left(2\pi f_{m}t-\phi\right)\right),
\eeq
where $f_{0}$ is the quasi-monochromatic frequency, $f_{m} = 1/T_{yr}$ is the LISA modulation frequency, $T_{yr}$ is the number of seconds in a year, $R_{\oplus}$ is the light travel time across 1 AU ($\sim$ 500 secs) and $(\theta,\phi)$ represent the sky location of the source.   The quantities $F^{+,\times}(t)$ are defined in the LFA by~\cite{rubbo}
\beq
F^{+}(t;\psi, \theta, \phi, \lambda) = \frac{1}{2}\left[\cos(2\psi)D^{+}(t;\theta, \phi, \lambda) - \sin(2\psi)D^{\times}(t;\theta, \phi, \lambda)\right],
\eeq
\beq
F^{\times}(t;\psi, \theta, \phi, \lambda) = \frac{1}{2}\left[\sin(2\psi)D^{+}(t;\theta, \phi, \lambda) + \cos(2\psi)D^{\times}(t;\theta, \phi, \lambda)\right],
\eeq
where $\psi$ is the GW polarization angle of the wave and $\lambda = 0$ or $3\pi/2$ defines the LISA A and E TDI channels.  The detector pattern functions $D^{+,\times}(t)$ are given by
\bea
D^{+}(t) = \frac{\sqrt{3}}{64}\left[\frac{}{}-36\sin^{2}(\theta)\sin(2\alpha(t)-2\lambda)+(3+\cos(2\theta)) \right.\\ \nonumber\fl \left(\frac{}{}\cos(2\phi)\left\{\frac{}{}9\sin(2\lambda)-\sin(4\alpha(t)-2\lambda)\right\} \frac{}{}+\sin(2\phi)\left\{\frac{}{}\cos\left(4\alpha(t)-2\lambda\right)-9\cos(2\lambda) \right\}\frac{}{}\right)\\ \nonumber  \left.-4\sqrt{3}\sin(2\theta)\left(\frac{}{}\sin(3\alpha(t)-2\lambda-\phi)-3\sin(\alpha(t)-2\lambda+\phi)\right)\right]
\eea
\bea
D^{\times}(t) = \frac{1}{16}\left[\frac{}{}\sqrt{3}\cos(\theta)\left(\frac{}{}9\cos(2\lambda-2\phi)-\cos(4\alpha(t)-2\lambda-2\phi) \right) \right. \\ \nonumber \left. -6\sin(\theta)\left(\frac{}{} \cos(3\alpha(t)-2\lambda-\phi)+3\cos(\alpha(t)-2\lambda+\phi) \right) \right],
\eea
where $\alpha(t)=2\pi t / T +\kappa$ is the orbital phase of the center of mass of the constellation and $\kappa$ is the initial ecliptic longitude, which we take to be zero for this study.

\subsection{The treatment of data gaps}
\begin{figure}[t]
\begin{center}
\epsfig{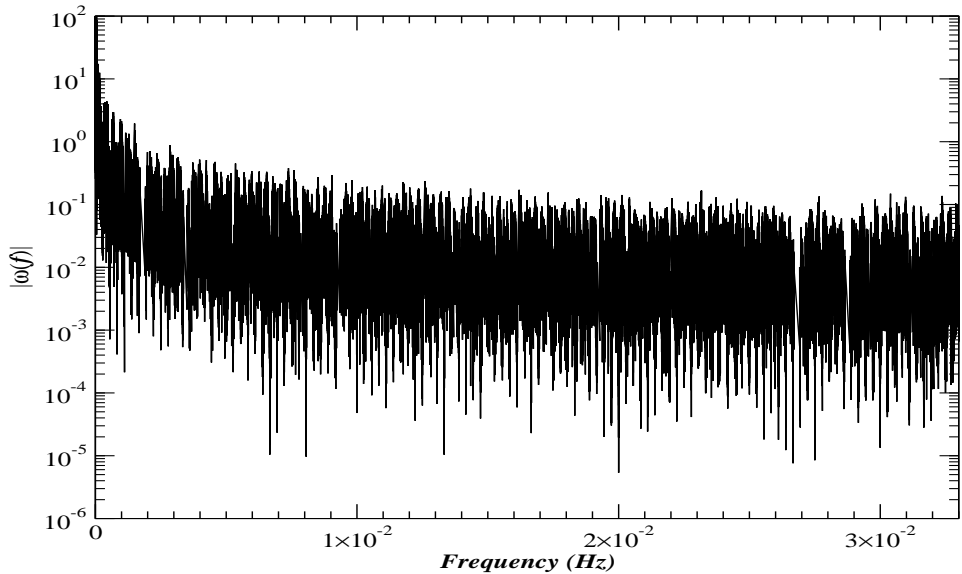}
\end{center}
\caption{Power spectrum of a response function composed of approximately one hour rectangular data gaps, once a day, over a period of one year.}
\label{fig:PSD_square}
\end{figure}

It is believed that there will be gaps in the LISA data stream due to a number of different possible phenomena.  While the 
frequency and duration of these gaps are at present unknown, we do know that we will need to somehow treat the gaps.  For practical purposes, we can consider
 the gaps in the data as the product between the LISA response $h(t)$, given by Equation (1), and a response function $\omega(t)$ consisting of a series of rectangular step functions, such that the gapped response is now $h_g(t)=h(t)\omega(t)$, where
 \beq
\omega(t) = \left\{ \begin{array}{ll} 0 & t_{gap}^i\leq t\leq t_{gap}^f \\ \\ 1 & t < t_{gap}^i, t > t_{gap}^f  \end{array}\right. .
\eeq
Here $t_{gap}^i$ and $t_{gap}^f$ are the time when the gap starts and finishes respectively.  If laser lock has been lost during the information down time, there
will be also be a transient response directly after the gap which would have a duration time on the order of minutes~\cite{Heinzel}.  Left untreated, a sudden discontinuity in the time domain data causes massive spectral leakage in the Fourier domain.  This is due to the fact that the Fourier transform of the rectangular function $\omega(t)$ is the sinc-function $\tilde{\omega}(f)=t\,sinc(\pi ft)$ which contains decaying frequency information as $f\rightarrow\infty$.  As an example, in Figure~\ref{fig:PSD_square} we plot the power spectrum of a response function that corresponds to an untreated data gap of approximately one hour, once a day, every day for a year.  We can see the high level of spectral leakage at higher frequencies in this response.   We also notice that after the first few sidelobes, there is a very slow drop-off in the amplitude of the lobes, which contribute spurious power to higher frequencies.  It is this excess power that we need to rid ourselves of as we search for GW sources.
 
\begin{figure}[t]
\begin{center}
\epsfig{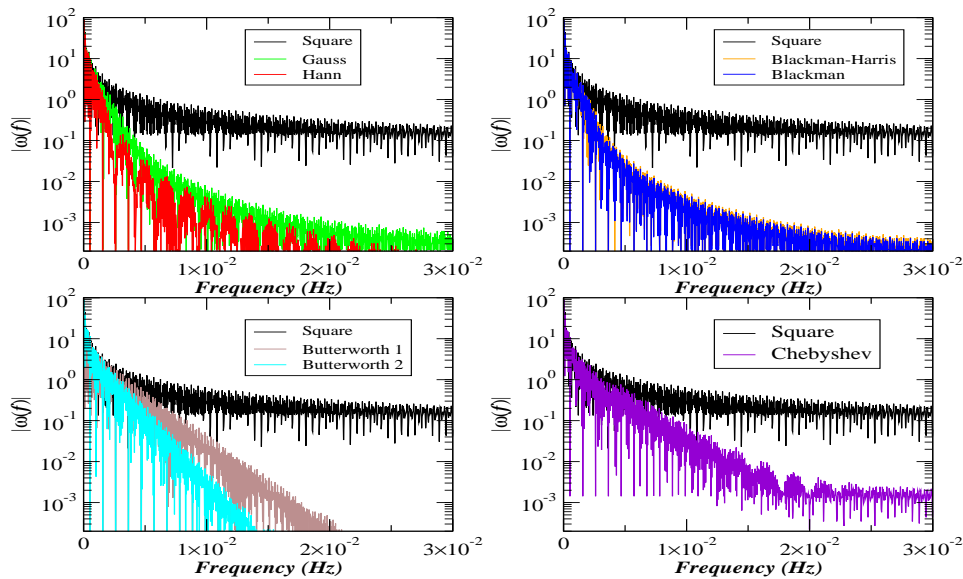}
\end{center}
\caption{A comparison of the power spectra of different window functions for a seven hour long duration, with a gap of one hour.  In each cell we also plot a rectangular response function as an example of the worst case function.}
\label{fig:PSD_windows}
\end{figure}

Before continuing, we should comment here on the option of interpolating the data gap.  It is common in many astronomical fields to try and interpolate the gapped data as short timescale data snippets may be taken over a long period of time (e.g. pulsar timing arrays).  The danger for LISA, we feel, is the large number of sources potentially in the data set.  If the data stream was composed solely of many millions of non-chirping galactic binaries, we could envisage a safe interpolation of the data as it would merely contain a superposition of overlapping sinusoids.  However, we could also imagine a more realistic scenario where, just before the gap, there is a supermassive black hole coalescence.  In order to interpolate the data, we need to use information from both sides of the gap.  The fact that we have a large finite signal ending just before the gap could massively distort the interpolation data leading to a spurious source detection in the gap.  We therefore have decided to err on the safe side of the ``\emph{garbage in, garbage out}" idiom, and not interpolate the data.  We will instead focusing on using a window function to smoothen the discontinuity and reduce the spectral leakage caused by the gaps in the time domain stream.

A perfect window function would have a narrow central peak (for high resolution), a low first sidelobe (indicating good noise suppression) and a rapid fall
off of sidelobes. However, in practice, this is difficult to achieve as filters with a narrow central
peak may have a slow fall off in sidelobes, while a window with rapid fall off in sidelobes may
have a wider central peak. Therefore, the goal is to strike a balance. We investigated a number of different window functions, as can be seen in Figure~\ref{fig:PSD_windows}, where we present the power spectra of a number of different windows. In each case the time series had a duration of seven hours with a one hour gap after approximately three hours.  In the end we chose the Hann function which strikes a nice balance between frequency resolution and noise suppression. For example, the rectangular window has its first zeros at $\pm 1$ bins, but the highest sidelobe is only -13.3 dB below the central peak and is located at 1.43 bins. Furthermore, the sidelobes fall off as $f^{-1}$. The Blackman-Harris window has a wider central peak with the first zero at $\pm 4$ bins, but with a lower first sidelobe, -92 dB below the central peak located at 4.52 bins. However, for this function, the sidelobes also fall off as $f^{-1}$. In contrast the Hann window has its first zero at $\pm 2$ bins. The highest sidelobe is at -31.5 dB and is located at $\pm2.36$ bins. Finally, the sidelobes fall off as $f^{-3}$. Therefore, with the Hann window we have an acceptable widening of the central peak, a low first sidelobe and the sidelobes fall off at a rapid rate. The Hann window used to treat the data gaps is defined as
 \beq
\omega(t) = \left\{ \begin{array}{ll} \frac{1}{2}\left(1-\cos\left(2\pi\frac{t-t_{gap}^i+2t_{win}}{2t_{win}}\right)\right) & t_{gap}^i-t_{win}< t< t_{gap}^i \\ \\ 0 & t_{gap}^i\leq t\leq t_{gap}^f \\ \\ \frac{1}{2}\left(1-\cos\left(2\pi\frac{t-t_{gap}^f}{2t_{win}}\right)\right) & t_{gap}^f < t< t_{gap}^f+t_{win}\\ \\1 & t \leq t_{gap}^i-t_{win},
 t \geq t_{gap}^f+t_{win}  \end{array}\right. ,
\eeq
where $t_{win}$ is the time necessary for the window function to go from 1 to 0, and conversely.  For this particular source, we experimented with a number of different values of $t_{win}$.  We found that $t_{win}=10$ minutes was sufficient in smoothening the rectangular window in order to reduce spectral leakage to acceptable levels.  


\section{The Monte Carlo Setup}
\label{Monte Carlo}

To evaluate the fidelity of one of our templates in the search for a GW signal, we define the noise-weighted inner product between a template $h(t)$ and a signal $s(t)$ as
\begin{equation}
\label{inner product}
\langle h \vert s \rangle = 2 \int_{0}^{\infty}\frac{\tilde{h}(f)\tilde{s}^{*}(f)+\tilde{h}^{*}(f)\tilde{s}(f)}{S_n(f)}df
\end{equation}
where $\tilde{h}(f)=\int_{-\infty}^{\infty}h(t)\exp (-2\pi ft)dt$ is the Fourier transform of $h(t)$, $\tilde{h}^{*}(f)$ is the complex conjugate of $\tilde{h}(f)$ and $S_n(f)=S_n^{instr}(f)+S_n^{conf}$ is the noise power spectral density of the LISA detector.  This quantity is the sum of instrumental noise and confusion noise from the background of unresolved galactic binaries. The instrumental noise for LISA is given by~\cite{cornish}
\bea
S_n^{instr}(f)=\frac{1}{4L^2}\left[2S_n^{pos}(f)\left(2+\cos^2\left(\frac{f}{f_*}\right)\right)\right. \\ \nonumber \left.+8S_n^{acc}(f)\left(1+\cos^2\left(\frac{f}{f_*}\right)\right)\left(\frac{1}{(2\pi f)^4}+\frac{\left(2\pi10^{-4}\right)^2}{(2\pi f)^6}\right)\right],
\eea
where $L=5\times10^6km$ is the LISA arm length, $S_n^{pos}(f)=4\times10^{-22}m^2Hz^{-1}$ and $S_n^{acc}(f)=9\times10^{-30}m^2s^{-4}Hz^{-1}$ are the position and acceleration noises respectively and $f_*=1/(2\pi L)$ is the mean transfer frequency for the LISA arm. For the galactic confusion noise, we use the following form derived from a Nelemans, Yungelson, Zwart galactic foreground model~\cite{NYP,trc}
 \beq
S_n^{conf}(f) = \left\{ \begin{array}{ll} 10^{-44.62}f^{-2.3} & 10^{-4}<f\leq 10^{-3}\\ \\ 10^{-50.92}f^{-4.4} & 10^{-3}<f\leq 10^{-2.7}\\ \\ 10^{-62.8}f^{-8.8} & 10^{-2.7}<f\leq 10^{-2.4}\\ \\ 10^{-89.68}f^{-20} & 10^{-2.4}<f\leq 10^{-2}\end{array}\right. ,
\eeq
where the confusion noise has units of $Hz^{-1}$.

Using the inner product in Equation~(\ref{inner product}), we define the SNR between signal and template as
\begin{equation}
\rho=\frac{\langle h \vert s \rangle}{\sqrt{\langle h \vert h \rangle}},
\end{equation}
and the overlap between the two waveforms in each channel as
\begin{equation}
O=\frac{\langle h \vert s \rangle}{\sqrt{\langle h \vert h \rangle \langle s \vert s \rangle}}.
\end{equation}
To conclude, we introduce a final tool, the Fisher information matrix (FIM) given by
\begin{equation}
F_{\mu\nu}=\left< \frac{\partial h}{\partial \lambda^{\mu}} \left| \frac{\partial h}{\partial \lambda^{\nu}} \right. \right>,
\end{equation}
the inverse of which is the variance-covariance matrix. In the limit of high SNR (which is almost always the case for LISA), the diagonal elements of the variance-covariance matrix provides an estimation of the variance in  the parameter estimation. In the case of monochromatic GBs, our system is defined by the parameter set $\lambda_{i}=\{ ln(A_0),i,\Psi,\varphi_0,ln(f_0),\cos(\theta),\phi \}$.

In order to quantify the effects of the data gaps on GB parameter estimation, we ran a full parameter Monte Carlo with 10,000 sources using a threshold of $\rho\geq5$. We chose this limit because it will be difficult at anything lower than this level to discern between a signal and noise. The duration and the frequency of planned gaps in the data are not known for the moment, so we choose a duration of one hour and we tested two gap frequencies : once per day and once per week. A gap duration of one hour was chosen as this should be sufficient to carry out operations on the spacecraft and regain laser lock, if lost during the disruption.  If the duration of all gaps is the same, and the gap frequency is constant during the observation, this adds (true) power to the two corresponding frequencies in the Fourier domain. To minimize this power we modulate the duration and the frequency with $\delta t_{dur}=\pm 15min$ and $\delta t_{freq}=\pm 5h$, respectively. As a monochromatic GB's signal is essentially constant over many years, the recovered SNR and parameter estimation are dependent on the duration of observation.  To test the effects of the gaps on both short and longer timescales, we ran MCs for each gap frequency assuming mission lifetimes of both one and three years.

The parameter limits for the MC were chosen such that the amplitude lies within the range $10^{-23}\leq A \leq 10^{-20}$, the monochromatic frequency of the binary was bounded by $10^{-4} \leq f_0/Hz \leq 5\times10^{-3}$ and all other parameters drawn from within their natural limits. In order to quantify the effects of the data gaps, we calculated the SNR and estimated parameter errors for cases with and without gaps. While the concept of a global overlap over both detector channels is a little ambiguous, it is possible to gain some idea of a goodness of fit by investigating the overlaps between the gapped and ungapped data in each channel.  


\section{Results}
\label{Results}

\begin{figure}[t]
\begin{center}
\epsfig{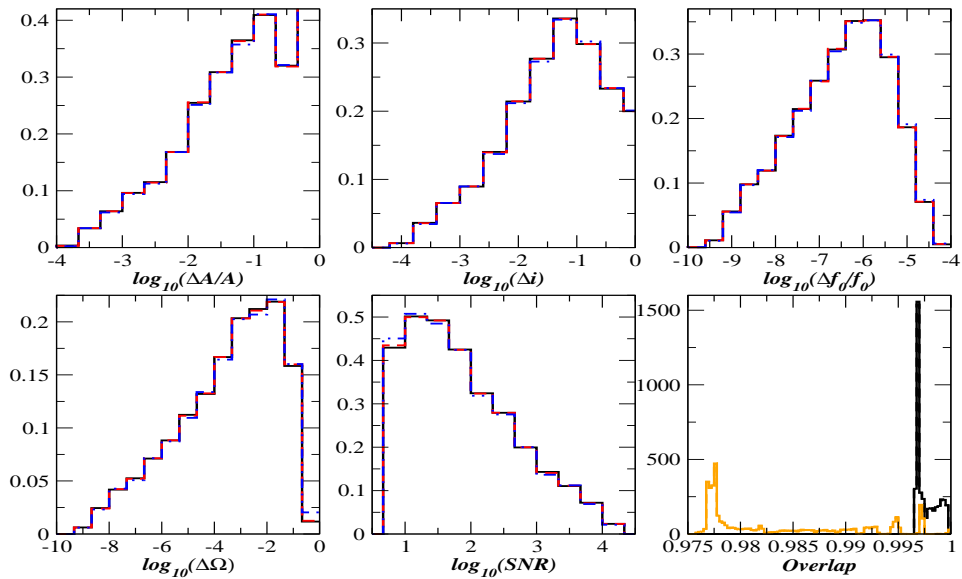}
\end{center}
\caption{Normalized histograms for distributions in the fractional error in the amplitude, the error in the inclination, the fractional error in the binary frequency, the sky resolution, the optimal SNR and overlaps between the sources, with and without gaps, for the LISA A and E channels assuming an observation time of one year. In the five first cells, going left to right, top to bottom,  we plot the results without gaps (black curve), with one gap per week (red dashed curve) and with one gap per day (blue dot-dashed curve). In the last cell we plot the overlaps for channels A and E for one gap per week (dark curves) and one gap per day (orange curves).  Note that the A and E channels are indistinguishable for each simulation.}
\label{fig:GB_1year}
\end{figure}

\begin{figure}[t]
\begin{center}
\epsfig{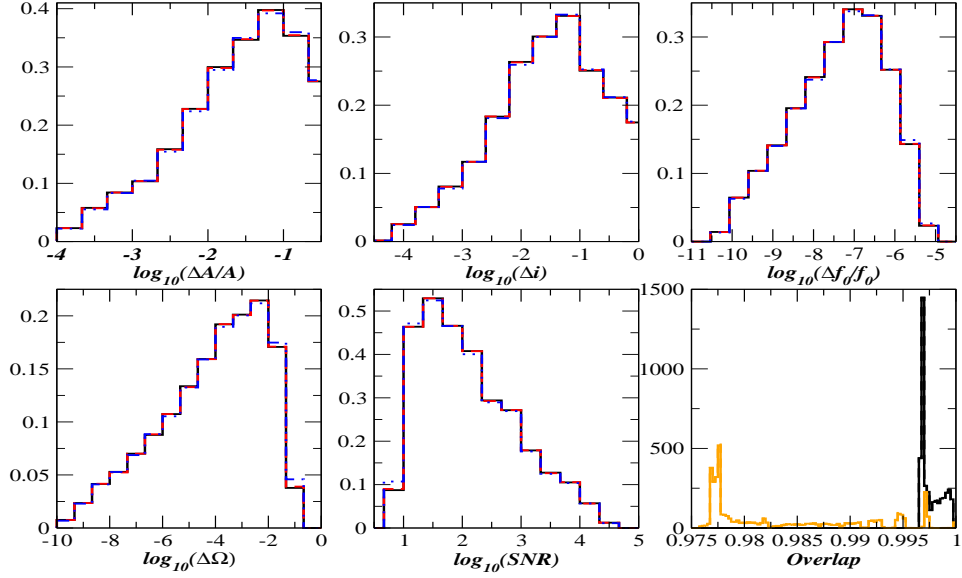}
\end{center}
\caption{Same as Figure~\ref{fig:GB_1year} but with an observation time of three years.}
\label{fig:GB_3years}
\end{figure}

We conducted MC simulations for observation times of both one and three year periods, and for two gap frequencies of once per day and once per week. In each run, for 10,000 sources with $\rho\geq 5$, we investigated recovered SNRs and parameter uncertainties with and without gaps.  We also ran a final case where we assumed a gap frequency of once per day over a period of one year, but with an added spurious data drop out that lasted a week in duration.  To quantify information loss due to the gaps,  we calculated overlaps between the waveform with gaps and the waveform without gaps in the A and E TDI channels.   In Figures~\ref{fig:GB_1year} and \ref{fig:GB_3years} we present our results for the two observation periods. For GBs, while the system is defined by a seven parameter set, only a subset are important for astronomical purposes.  Therefore, each of the graphs are organised in the following way. On the top row, we present the fractional error in the amplitude, the error in the inclination of the binary plane and the fractional error in the binary frequency. On the bottom row, we present the sky resolution $\Delta \Omega = 2\pi \sqrt{ \Sigma^{\theta\theta} \Sigma^{\phi\phi} - \left(\Sigma^{\theta\phi}\right)^2}$ (where $\Sigma^{ij}$ are the components of the inverse FIM and the sky resolution has units of steradians), the optimal SNR and finally, the overlaps between the sources with and without gaps, for the LISA A and E channels. In each cell describing the parameter errors and SNR, the black curve corresponds to no data gaps, the red dashed line denotes the case of one gap per week and blue dot-dashed line represents the case of one gap per week.  In the final cell, the dark lines (solid and dashed) represent the overlaps between gapped and ungapped data in the case of one gap per week, while the orange lines (solid and dashed) denote the case of one gap per day.

The first thing that is observable from both figures is that it is very difficult to visually make out a difference between the three cases for the estimation of parameters and recovered SNRs.  What is obvious is that, for both one and three year observation periods, the median overlap between a gapped template and an ungapped template drops from almost 99\% in both channels for one gap a week, to almost 97\% for one gap a day.  This demonstrates that the more frequent gaps do have an effect on parameter estimation which is not picked up visually in the other cells. 

In Table 1 we present the median error prediction in the estimation of parameters for both the one and three year observation periods for all data gap frequencies.  In the case
of a one year observation period with a gap of one hour, once a week, the effect of the data gaps is the introduction of a bias of between 0.5-1\% error in the estimation of the amplitude, frequency and sky resolution.  Once we go to one gap per day, this error increases to between 3-7\% depending on the parameter.  The inclusion of a spurious one
week gap increased the bias to between 4-9\% for the important parameters.  The decrease in recovered SNR follows a similar pattern.  Once we go to the three year mission case, the situation does not change with the estimated biases remaining almost constant.  This is not really a surprising result as the monochromatic binary essentially outputs
a constant power during the mission lifetime.  The effect of the data gaps is the elimination of a small number of data points per observation period.  This means that the total number of missing data points scales with, and is compensated for, by the observation period.  We do not expect this to be the case for sources with significant frequency evolution, as more and more data points are lost within each subsequent gap.

\begin{table}
\centering
\begin{tabular}{c c c c c}
\hline
Parameter 			& 1 year 					& 1 year					 & 1 year							& 1 year\\
 				& No gaps					& 1 / week					& 1 / day							& 1 / day + 1 week\\
\hline
$\Delta A/A$ 		& $1.165803 \times 10^{-1}$	& $1.172101 \times 10^{-1}$	& $1.210735 \times 10^{-1}$		& $1.222675 \times 10^{-1}$\\
$\Delta \iota$/rad		& $1.278432 \times 10^{-1}$	& $1.284081 \times 10^{-1}$ 	& $1.320791 \times 10^{-1}$		& $1.333090 \times 10^{-1}$\\
$\Delta \Psi$ /rad		& $1.877202 \times 10^{-1}$ 	& $1.885523 \times 10^{-1}$ 	& $1.944390 \times 10^{-1}$		& $1.966201 \times 10^{-1}$\\
$\Delta \varphi_{0}$/rad 	& $4.220753 \times 10^{-1}$ 	& $4.228617 \times 10^{-1}$ 	& $4.339971 \times 10^{-1}$		& $4.384876 \times 10^{-1}$\\
$\Delta f_{0}/f_{0}$	& $4.122052 \times 10^{-7}$ & $4.139029 \times 10^{-7}$ & $4.250480 \times 10^{-7}$ & $4.289691 \times 10^{-7}$\\
$\Delta \Omega$/ster & $7.379220 \times 10^{-4}$ & $7.452740 \times 10^{-4}$ & $7.861529 \times 10^{-4}$ & $8.009114 \times 10^{-4}$\\
$SNR$	 & $53.27678$ & $52.97586$ & $51.49511$ & $50.85469$\\
$O_{A}$	 & $1$ & $0.997110$ & $0.979505$ & $0.973763$\\
$O_{E}$	 & $1$ & $0.997109$ & $0.979511$ & $0.973634$\\
\hline \hline
Parameter 			& 3 years 					& 3 years 					& 3 years\\
		 		& No gaps 					& 1 / week 				& 1 / day\\
\hline
$\Delta A/A$ 		& $6.711915 \times 10^{-2}$ 	& $6.752622 \times 10^{-2}$ 	& $6.973013 \times 10^{-2}$\\
$\Delta \iota$/rad		& $7.366606 \times 10^{-2}$ 	& $7.401683 \times 10^{-2}$ 	& $7.613257 \times 10^{-2}$\\
$\Delta \Psi$/rad 		& $1.012134 \times 10^{-1}$ 	& $1.016863 \times 10^{-1}$ 	& $1.048081 \times 10^{-1}$\\
$\Delta \varphi_{0}$/rad 	& $2.128610 \times 10^{-1}$ 	& $2.137969 \times 10^{-1}$ 	& $2.204804 \times 10^{-1}$\\
$\Delta f_{0}/f_{0}$	& $5.713411 \times 10^{-8}$ 	& $5.737847 \times 10^{-8}$ 	& $5.898288 \times 10^{-8}$\\
$\Delta \Omega$/ster 	& $1.796017 \times 10^{-4}$ 	& $1.810762 \times 10^{-4}$ 	& $1.915378 \times 10^{-4}$\\
$SNR$			& $92.38947$ 				& $91.90973$ 				& $89.25755$\\
$O_{A}$			& $1$	 				& $0.997093$ 				& $0.979487$\\
$O_{E}$			& $1$	 				& $0.997093$				& $0.979488$\\
\hline
\end{tabular}
\caption{Median parameter uncertainties, SNRs and single channel overlaps for galactic binaries assuming different data gap frequencies, and mission timescales of one and three years.}
\label{table:GB_parameter_uncertainties}
\end{table}


\section{Conclusions}\label{sec:conclusions}
In this study, we have investigated the effects of data gaps on the estimation of parameters for galactic binaries with LISA.  We have
demonstrated that left untreated, the data gaps cause massive spectral leakage in the Fourier domain, where the majority of the data
analysis is conducted.  This spectral leakage would have a large effect on the estimation of parameters and could in fact lead to spurious
detections.  Using a Hann window function, we smoothen the discontinuities in the time-domain data stream in order to reduce the leakage
as best as possible.  

In order to prove the validity of our window function, we then ran a $10^5$ iteration Monte Carlo simulation to 
investigate the estimation of parameter errors using a complete data set, a gapped but untreated data set and finally and gapped and 
treated data set for both one and three year LISA missions.  As the frequency and duration of the gaps are presently unknown, we simulated
scenarios where we have one gap per day and one gap per week, where each gap has a $60\pm15$ minute duration.  We found that a gap
of once per week introduces a bias of between 0.5\% and 1\% in the parameter estimation and allows us to recover more than a 99\% correlation with a perfect data set containing no gaps.  However, if we have to contend with one gap per day, the correlation 
with a perfect data set drops to about 97\% and the parameter error increases by 3\% to 7\% depending on the parameter being investigated.  We finally simulated a scenario where, as well as the gap of
once per day, we also had a spurious week long data gap.  In this particular case the bias in parameter estimation increases to between 4\% and 9\% for the most important parameters.   The overwhelming conclusion for galactic binaries, is that in order to reduce any bias in the parameter estimation due to the gaps to under 1\%, we require a gap frequency of no more than once per week.  In the future we plan to conduct an investigation of the effect of gaps on 
sources such as the inspirals of supermassive black hole binaries and EMRIs where we feel the effect of the gaps may be more pronounced.

\section*{Acknowledgements}
The authors would like to thank the following people for useful conversations and suggestions : P. Bender, O. Jennrich, R. Stebbins, G. Heinzel, E. Plagnol, G. Auger and H. Halloin.

\end{document}